\documentstyle[aps, pra, twocolumn]{revtex}
\begin{document}
\draft
\begin{title}
{Gap and subgap tunnelling in  cuprates }
\end{title} 
\author{A.S. Alexandrov$^*$ and A.F. Andreev$^{**}$}
\address
{* Department of Physics, Loughborough University, Loughborough LE11 
3TU, United Kingdom, **P.L. Kapitza Institute for Physical Problems, Kosygin
st.2, 117973- Moscow, GSP-1,  Russia }

\maketitle
\begin{abstract}
We  describe strongly attractive carriers in  
cuprates in the framework of a simple quasi-one dimensional
Hamiltonian with a local attraction.
 In contrast with  the conventional BCS theory
there are two  energy scales, a temperature independent 
 incoherent gap $\Delta_p$ and a temperature dependent 
coherent gap  $\Delta_c (T)$ combining into one temperature
dependent global gap $\Delta=(\Delta_p^2 +\Delta_c^2)^{1/2}$. 
  The temperature dependence of the 
  gap and single particle (Giaver) tunnelling spectra  in cuprates are
  quantitatively described. A framework for understanding of two distinct energy scales
  observed  in Giaver tunnelling and electron-hole reflection
experiments is provided.
\end{abstract}
\narrowtext

 There is  convincing experimental evidence that  the  pairing of carriers takes
 place well above T$_c$ in underdoped cuprates (for a review see Ref.
 \cite{alemot}).  
If carriers are paired their magnetic moments compensate each other
 so one could  expect 
that the normal state uniform  magnetization should fall  with decreasing 
temperature  because more and more holes are bound into singlet pairs. This 
 unexpected drop of the normal state magnetic susceptibility  
 was experimentally observed \cite{jon,mul} and explained 
   in the framework of the bipolaron theory of cuprates
 \cite{alekab,mul}.
There is also  a  gap in the tunnelling and  photoemission, which is almost temperature independent below T$_{c}$ \cite{brus1} and  exists well above
T$_{c}$ \cite{ren,brus2,she}, so that some  segments of a `large Fermi
 surface' are actually missing \cite{bia,bia2}.      Kinetic
\cite{bat} and  thermodynamic \cite{lor} data suggest that the gap
 opens in both charge and spin channels and exists at any relevant
temperature in a wide range of doping. A plausible explanation is that
 the normal gap is half of the bipolaron binding energy \cite{alemot},
 although alternative explanations have also been proposed. The  temperature and doping
dependence of the gap still remains a subject of controversy. Moreover, 
reflection experiments, in which an incoming electron from the normal
 side of a normal/superconducting contact is reflected as a hole along
 the same trajectory \cite{and}, revealed a much smaller  gap edge than the
 bias
at the tunnelling conductance maxima in a few  underdoped cuprates
 \cite{yag}.  Recent intrinsic tunnelling measurements on a series of
 Bi '2212' single crystals \cite{pmul} showed distinctly different
 behaviour of the superconducting and normal state gaps with the
 magnetic field. Such coexistance of two distinct gaps in cuprates
 is not well understood \cite{deu2,pmul}.

In this letter we propose a model, which describes the
temperature dependence of the gap, tunnelling spectra and electron-hole reflection
in cuprates. The  assumption is that the attraction potential in
cuprates is large compared with  the Fermi energy. The main point of our letter is  independent of
 the microscopic nature of the attraction.  Real-space  pairs  might be
lattice and/or spin bipolarons \cite{alemot}, or any other preformed pairs.

We start with  a  generic one-dimensional  Hamiltonian including  the
kinetic energy of carriers in the effective mass ($m$) approximation and
a  local attraction potential, $V(x-x')=-U \delta(x-x')$ as
\begin{eqnarray}
 H &=& \sum_{s}\int dx \psi_s^{\dagger}(x)
 \left(-{1\over{2m}}{d^2\over{dx^2}} -\mu \right) \psi_s(x)\cr
&-& U \int dx \psi_\uparrow^{\dagger}(x)\psi_\downarrow^{\dagger}(x)\psi_\downarrow(x)\psi_\uparrow(x),
 \end{eqnarray}
 where $s= \uparrow,\downarrow$ is the 
 spin ($\hbar=k_{B}=1$). The first band to be doped in cuprates is the
 oxygen band inside the Hubbard gap as established in  polarised
 photoemission  \cite{mat,aleden}. This band is almost one
 dimensional as discussed in Ref. \cite{ale}, so that our
(quasi) one-dimensional approximation is a realistic starting point.

Solving a two-particle problem with the $\delta$-function potential
one obtains a bound state with the binding energy
\begin{equation}
2\Delta_{p}= {1\over{4}}mU^2,
\end{equation}
and with the radius of the bound state $r=2/mU$. We assume that this radius is less
than the inter-carrier distance in  cuprates, which puts a
constraint on the  doping level, $E_F < 2\Delta_p$, where $E_F$ is the free-carrier
Fermi energy.  Then  real-space  pairs are formed. If   three-dimensional corrections to the
energy spectrum of pairs are taken into account (see, for example,
Ref. \cite{alekab2}) 
 the ground state of the system is the Bose-Einstein condensate.  The
 chemical potential is pinned
 below the band edge by about  $\Delta_p$  both in the superconducting
 and normal state \cite{alemot}, so that the normal state
single-particle gap is $\Delta_p$. The binding energy $2\Delta_p$
might change due to the same corrections. However,  this change does not
affect  our further results as soon as they are expressed in terms
of $\Delta_p$ rather than $U$. 

Now  we take into
account that in the superconducting state ($T<T_c$)
the single-particle excitations
 interact with the condensate via the same potential $U$. Applying the
 Bogoliubov approximation \cite{bog} we reduce the Hamiltonian, Eq.(1) to a
 quadratic form as
\begin{eqnarray}
 H &=& \sum_{s}\int dx \psi_s^{\dagger}(x)
 \left(-{1\over{2m}}{d^2\over{dx^2}} -\mu \right) \psi_s(x)\cr
&+&\int dx [\Delta_c \psi_\uparrow^{\dagger}(x)\psi_\downarrow^{\dagger}(x)+H.c.],
 \end{eqnarray}
where the coherent pairing potential
\begin{equation}
\Delta_c=-U\langle \psi_\downarrow(x)\psi_\uparrow(x)\rangle
\end{equation}
is proportional to the square root of the condensate density,
$\Delta_c =constant \times  n_0(T)^{1/2}$.  The single-particle excitation energy
spectrum $E(k)$ is found using the Bogoliubov transformation as
\begin{equation}
E(k)= \left[(k^2/2m+\Delta_p)^2+\Delta_c^2\right]^{1/2},
\end{equation}
if one assumes  that the condensate density does not depend on position. This spectrum is quite different
from the BCS quasiparticles because the chemical potential is negative,
$\mu=-\Delta_p$. The single particle gap, $\Delta$, defined as  the minimum of
$E(k)$, is given  by
\begin{equation}
\Delta= \left[\Delta_p^2+\Delta_c^2 \right]^{1/2}.
\end{equation}
It varies with temperature from $\Delta(0)=
\left[\Delta_p^2+\Delta_c(0)^2 \right]^{1/2}$ at zero temperature down to
the temperature independent $\Delta_p$  above $T_c$. The condensate
density depends on temperature as $1-(T/T_c)^{d/2}$ in the ideal
 three ($d=3$) and (quasi) two-dimensional ($d=2$) Bose-gas.  In the three-dimensional $charged$ Bose-gas it has an
exponential temperature dependence at low temperatures due to a plasma
gap  in the Bogoliubov collective excitation spectrum \cite{fol}, which might
be highly anisotropic in cuprates \cite{alemot}. Near T$_c$ 
one can expect a power law dependence, $n_0(T) \propto 1- (T/T_c)^n$
with $n>d/2$  because the condensate plasmon \cite{fol}
depends on temperature.
The theoretical temperature dependence, Eq.(6) describes well  the pioneering experimental observation of
the anomalous gap in $YBa_2Cu_3O_{7-\delta}$ in the
electron-energy-loss spectra by Demuth $et$ $al$ \cite{dem}, Fig.1, with
$\Delta_c(T)^2=\Delta_c(0)^2 \times[1-(T/T_c)^n]$ below $T_c$ and 
 zero above $T_c$, and $n=4$.

 The normal metal-superconductor (SIN)  tunnelling conductance  via a dielectric contact, $dI/dV$ is proportional
 to the density of states, $\rho(E)$  of the spectrum 
 Eq.(5).  Taking also into account the scattering of single-particle excitations
 by a random potential, thermal lattice and
 spin fluctuations one  finds at $T=0$\cite{ale}
\begin{equation}
dI/dV=constant \times   
[\rho\left({2eV-2\Delta\over{\epsilon_{0}}}\right)+
A\rho\left({-2eV-2\Delta\over{\epsilon_{0}}}\right)],
\end{equation}
with
\begin{equation}
\rho(\xi)={4\over{\pi^{2}}}\times{Ai(-2\xi)
Ai'(-2\xi)+
Bi(-2\xi)Bi'(-2\xi)\over{
[Ai(-2\xi)^{2}
+Bi(-2\xi)^{2}]^{2}}},
\end{equation}
$A$ is the asymmetry coefficient\cite{ale}, $Ai(x), Bi(x)$ the Airy functions, and $\epsilon_0$ is the
scattering rate. 
We compare the  conductance, Eq.(7) with  one of the best
 STM
 spectra measured in $Ni$-substituted $Bi_2Sr_2CaCu_2O_{8+x}$ single
crystals by Hancottee $et$ $al $\cite{brus1}, Fig.2. This experiment showed anomalously
large $2\Delta/T_c > 12$ with the temperature dependence of the
gap similar to that in Fig.1.  
 The theoretical conductance, Eq.(7) describes well 
 the 
anomalous $gap/T_{c}$ ratio, injection/emission assymmetry,
zero-bias conductance at zero temperature,  and the spectral shape inside and 
outside the gap region. There is no doubt that the gap, Fig.2 is
s-like, which is compatible with the phase-sensitive experiments
\cite{pha} in the framework of the bipolaron theory\cite{ale}.
Within the theory the single-particle gap might be almost $k$
independent while the symmetry of the Bose-Einstein condensate
wave-function (i.e. of the order parameter) is
 $d-wave$.

Finally, we propose a simple theory of the tunnelling into bosonic
 (bipolaronic) superconductor in the metallic (no-barrier)
regime. As in the canonical BCS approach applied to the normal
 metal-superconductor tunnelling  by Blonder, Tinkham and
Klapwijk \cite{btk} and to the normal-superconductor boundary in the
 intermediate type I state  by one of us \cite{and}, the incoming electron produces
only outgoing particles in the superconductor ($x>l$), allowing for a
 reflected electron and (Andreev) hole in the normal
metal ($x<0$). There is also a buffer layer of the thickness $l$ at the normal
metal-superconductor boundary ( $x=0$), where the chemical potential 
with respect to the bottom of the conduction band changes gradually
from a positive large value $\mu$ in the metal to a negative 
value $-\Delta_p$ in the bosonic superconductor. We approximate this buffer layer  by 
 a layer with a constant chemical potential $\mu_b$ ($-\Delta_p<\mu_{b}<\mu$) and with the
 same strength of the pairing potential $\Delta_c$ as in the bulk
 superconductor.  The Bogoliubov-de Gennes equations may be written
as usual  \cite{btk}, with the only difference that the chemical
 potential with respect to the bottom of the band is a function of the
 coordinate $x$,
\begin{eqnarray}
\left 
(\matrix{-(1/2m) d^2/dx^2 -\mu(x)& \Delta_c \cr \Delta_c &(1/2m) d^2/dx^2 +\mu(x) }\right) \psi(x)\cr
=E\psi(x).
\end{eqnarray}
  Thus the two-componet wave function in the normal
 metal is given by  
\begin{equation}
\psi_n(x<0)=\left 
(\matrix{1 \cr 0 }\right)e^{iq^+x}+b\left 
(\matrix{1 \cr 0 }\right)e^{-iq^+x}+a\left 
(\matrix{0\cr 1 }\right)e^{-iq^-x},
\end{equation}
while in the buffer layer it has the form
\begin{eqnarray}
\psi_b(0<x<l)&=&\alpha \left 
(\matrix{1 \cr {\Delta_c\over{E+\xi}} }\right)e^{ip^+x}+\beta \left 
(\matrix{1 \cr  {\Delta_c\over{E-\xi}}}\right)e^{-ip^-x}\cr
&+&\gamma\left 
(\matrix{1\cr  {\Delta_c\over{E+\xi}} }\right)e^{-ip^+x}+\delta \left 
(\matrix{1 \cr  {\Delta_c\over{E-\xi}}}\right)e^{ip^-x},
\end{eqnarray}
where  the momenta associated with the energy $E$ are
\begin{equation}
q^{\pm}=[2m(\mu \pm E )]^{1/2}
\end{equation}
and
\begin{equation}
p^{\pm}=[2m(\mu_b \pm \xi )]^{1/2}
\end{equation}
with $\xi= (E^2-\Delta_c^2)^{1/2}$. The well-behaved solution in the
superconductor with  negative chemical potential is given by
\begin{equation}
\psi_s(x>l)=c \left 
(\matrix{1 \cr  {\Delta_c\over{E+\xi}} }\right)e^{ik^+x}+d \left 
(\matrix{1 \cr {\Delta_c\over{E-\xi}}   }\right)e^{ik^-x},
\end{equation}
where  the momenta associated with the energy $E$ are
\begin{equation}
k^{\pm}=[2m(-\Delta_p \pm \xi )]^{1/2}.
\end{equation}
The coefficients $a,b,c,d, \alpha,\beta,\gamma,\delta$ are determined
from the boundary conditions, which are continuity of $\psi(x)$ and its
derivatives at $x=0$ and  $x=l$. Applying the boundary conditions,
and carrying out an algebraic reduction, we find
\begin{equation}
a= 2\Delta_c q^+(p^+f^-g^+ -p^-f^+g^-)/D,
\end{equation}
\begin{eqnarray}
b&=&-1 +2q^+[(E+\xi)f^+(q^-f^--p^-g^-)\cr
&-&(E-\xi)f^-(q^-f^+-p^+g^+)]/D,
\end{eqnarray}
with
\begin{eqnarray}
D&=&(E+\xi)(q^+f^++p^+g^+)(q^-f^--p^-g^-)\cr
&-&(E-\xi)(q^+f^-+p^-g^-)(q^-f^+-p^+g^+),
\end{eqnarray}
and $f^\pm= p^\pm \cos(p^\pm l)-ik^\pm \sin(p^\pm l)$, $g^\pm=k^\pm
\cos(p^\pm l) -i p^\pm \sin(p^\pm l)$.

The transmisson coefficient for electrical current,
$1+|a|^2-|b|^2$ is shown in Fig.3 
for different values of $l$ when the coherent gap $\Delta_c$ is
smaller than  the pair-breaking gap $\Delta_p$, and in Fig.4 for the
opposite case, $\Delta_p<\Delta_c$. In the first case, Fig.3, we find two
distinct energy scales, one is $\Delta_c$ in the subgap region due to 
electron-hole  reflection and the other one is $\Delta$, which is the
single-particle band edge. On the other hand,  there is only one gap $\Delta_c$, which can
be seen in the second case, Fig.4. We notice that the transmission has no
subgap structure if the buffer layer is absent ($l=0$) in both cases. In the
extreme case of a wide buffer layer, $l>> (2m\Delta_p)^{-1/2}$, Fig.3,
or $l>>(2m\Delta_c)^{-1/2}$, Fig.4, there are some oscillations of the
transmission due to the bound states inside the buffer layer. It was
shown in Ref. \cite{alekab} that the pair-breaking gap $\Delta_p$  is inverse
proportional to the doping level. On the other hand, the coherent gap $\Delta_c$ 
scales with the condensate density, and therefore with the critical
temperature, determined as the Bose-Einstein condensation temperature
of strongly anisotropic 3D bosons \cite{alekab2}.
Therefore we expect that $\Delta_p>>\Delta_c$ in the underdoped
cuprates, Fig.3, while $\Delta_p \leq \Delta_c$ in the optimally doped
cuprates, Fig.4. Thus the model accounts for the two different gaps experimentally
observed in  Giaver tunnelling and electron-hole reflection in the
underdoped cuprates and for a single gap in the optimally doped
samples \cite{deu2}. The transmission,
Fig.3 and Fig.4, is entirely due to the coherent tunnelling into the
condensate and (or) into the single-particle band of the  bosonic
superconductor. There is also an incoherent transmission into localised
single-particle  impurity
states and into incoherent ('supracondensate') bound pair states,
 which might explain a significant  featureless  background
in the subgap region \cite{yag}.

In conclusion, we have proposed a simple  general model, which
 provides an explanation of the temperature dependence of the gap and of
the single-particle tunnelling spectra in cuprates. The main assumption is
 that the attractive potential is large compared with the Fermi
 energy, so that the ground state is the Bose-Einstein condensate of
 tightly bound pairs. We have developed a theory of  tunnelling in
 the metallic regime with no barrier and  found two
 different energy scales in the transmission as observed experimentally.

We acknowledge support of this work  by the Leverhulme Trust (London),
grant VP/261.

\centerline{{\bf Figure Captures}}

Fig.1. Temperature dependence of the gap, Eq.(6) (line) compared with 
the experiment \cite{dem}(dots) for $\Delta_p=0.7 \Delta(0)$ .

Fig.2. Theoretical tunnelling conductance, Eq.(7)  (line) compared with   STM 
conductance 
in Ni-doped $Bi_2Sr_2CaCu_2O_{8+x}$ \cite{brus1}) (dots) for $2\Delta=90$ meV,
$A=1.05$, $\epsilon_0=40$ meV.

Fig.3. Transmission versus voltage (measured in units of  $\Delta_p/e$)
for $\Delta_c=0.2\Delta_p$, $\mu=10 \Delta_p$, $\mu_b=2 \Delta_p$ and  
$l=0$ (thick  line), $l=1$ (thick dashed line), $l=4$ (thin line), and
$l=8$ (thin dashed line) (in units of $1/(2m \Delta_p)^{1/2}$).

Fig.4. Transmission versus voltage (measured in units of $\Delta_c/e$) for
$\Delta_p=0.2 \Delta_c$, $\mu=10 \Delta_c$, $\mu_b=2\Delta_c$ and 
$l=0$ (thick  line), $l=1$ (thick dashed line), $l=4$ (thin line), and
$l=8$ (thin dashed line) (in units of $1/(2m \Delta_c)^{1/2}$).


\begin{thebibliography}{99}


\bibitem{alemot}
A.S. Alexandrov and N.F. Mott, Rep. Prog. Phys. {\bf 57} 1197 (1994).
\bibitem{jon}
D.C. Johnston, Phys. Rev. Lett {\bf 62}, 957 (1989).
\bibitem{mul}
K.A. M\"uller $et$ $al$, J.Phys.: Condens. Matter {\bf 10}, L291 
(1998).
\bibitem{alekab}
A.S. Alexandrov, V.V. Kabanov and N.F. Mott, Phys. Rev. Lett. {\bf 77}, 4796 (1996)
\bibitem{brus1}
H. Hancotte $et$ $al$, Phys. Rev. B{\bf 55}, R3410 (1997).
\bibitem{ren}
Ch. Renner $et$ $al$, Phys. Rev. Lett. {\bf 80}, 149 (1998).
\bibitem{brus2}
A. Mourachkine, Europhys. Lett. {\bf 49}, 86 (2000).
\bibitem{she}
Z.-X. Shen and J.R. Schrieffer, Phys. Rev. Lett. {\bf 78}, 1771
(1997) and references therein.
\bibitem{bia}
N.L. Saini $et$ $al$, Phys. Rev. Lett. {\bf 79}, 3467 (1997).
\bibitem{bia2}
N.L. Saini $et$ $al$, Phys. Rev. Lett. {\bf 82}, 2619 (1999)
\bibitem{bat}
B. Batlogg $et$ $al$, Physica C (Amsterdam) {\bf 135-140}, 130 (1994);
\bibitem{lor}
J.W. Loram $et$ $al$, Physica C (Amsterdam), {\bf 235}, 134 (1994).
\bibitem{and}
A.F. Andreev, Zh. Eksp. Teor. Fiz. {\bf 46}, 1823 (1964) [Sov.
Phys. JETP {\bf 19}, 1228 (1964)].
\bibitem{yag}
Y. Yagil $et$ $al$, Physica C (Amsterdam) {\bf 250}, 59 (1995).
\bibitem{pmul}
P. M\"uller $et$ $al$, unpublished.
\bibitem{deu2}
G. Deutscher, Nature {\bf 397}, 410 (1999).
\bibitem{mat}
M. C. Schabel $et$ $al$, Phys. Rev. B {\bf 57}, 6090 (1998).
\bibitem{aleden}
A.S. Alexandrov and C.J. Dent, Phys. Rev. {\bf
  60}, 15414 (1999).
\bibitem{ale}
A.S. Alexandrov, Physica C (Amsterdam) {\bf 305}, 46 (1998).
\bibitem{alekab2}
A.S. Alexandrov and V.V. Kabanov, Phys. Rev. B {\bf 59}, 13628 (1999).
\bibitem{bog}
N. Bogoliubov, J.Phys. USSR {\bf 11}, 23-32(1947).
\bibitem{fol}
L.L. Foldy, Phys.Rev. {\bf 124}, 649(1961).
\bibitem{dem}
J.E. Demuth $et$ $al$, Phys. Rev. Lett. {\bf 64}, 603 (1990)
\bibitem{pha}
D.A. Wollman $et$ $al$, Phys. Rev. Lett. {\bf 71}, 2134 (1993); C.C.
Tsuei $et$ $al$, Phys. Rev. Lett. {\bf 73}, 593 (1994); J.R. Kirtley
$et$ $al$, Nature {\bf 373}, 225 (1995); C.C. Tsuei $et$ $al$,
Science {\bf 272}, 329 (1996).
\bibitem{btk}
G.E. Blonder, M. Tinkham, and T.M. Klapwijk, Phys. Rev. B{\bf 25},
4515 (1982).







\end{thebibliography}
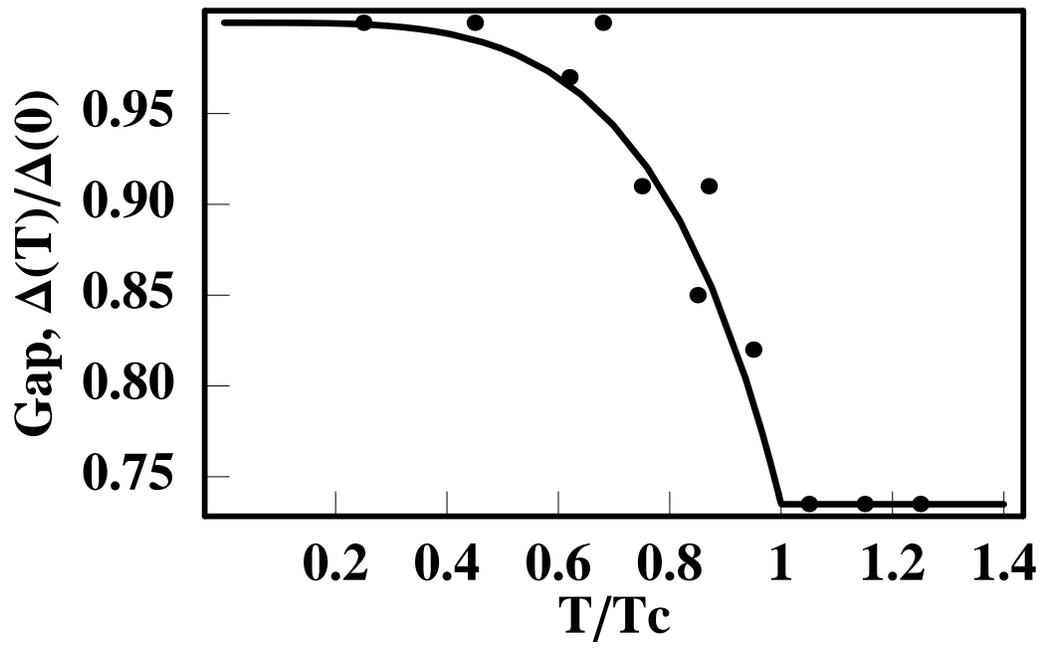
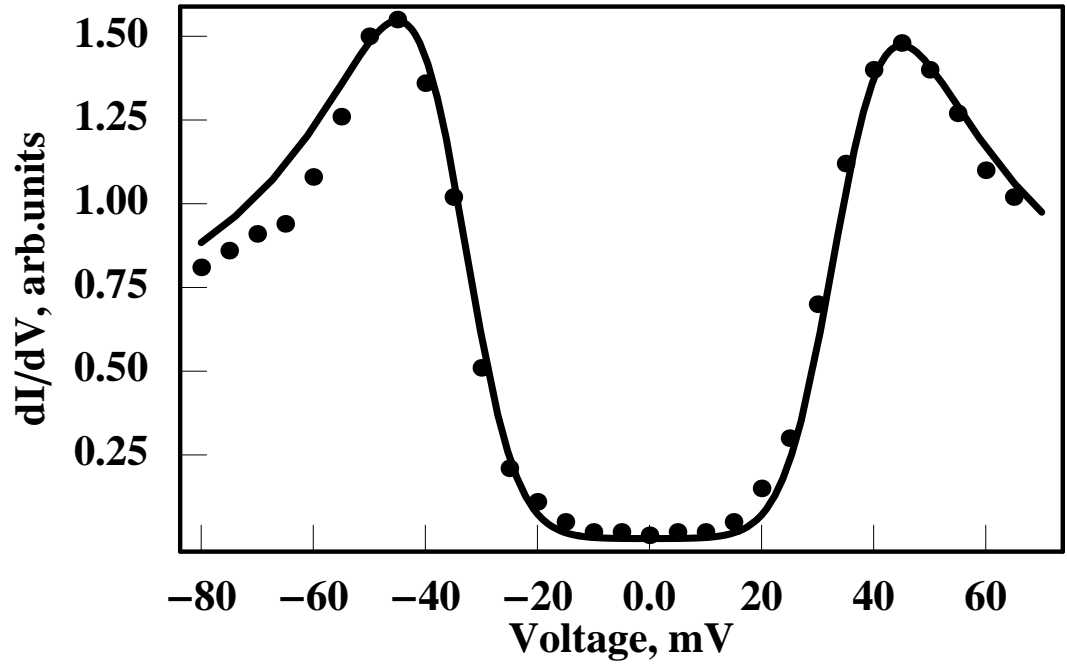
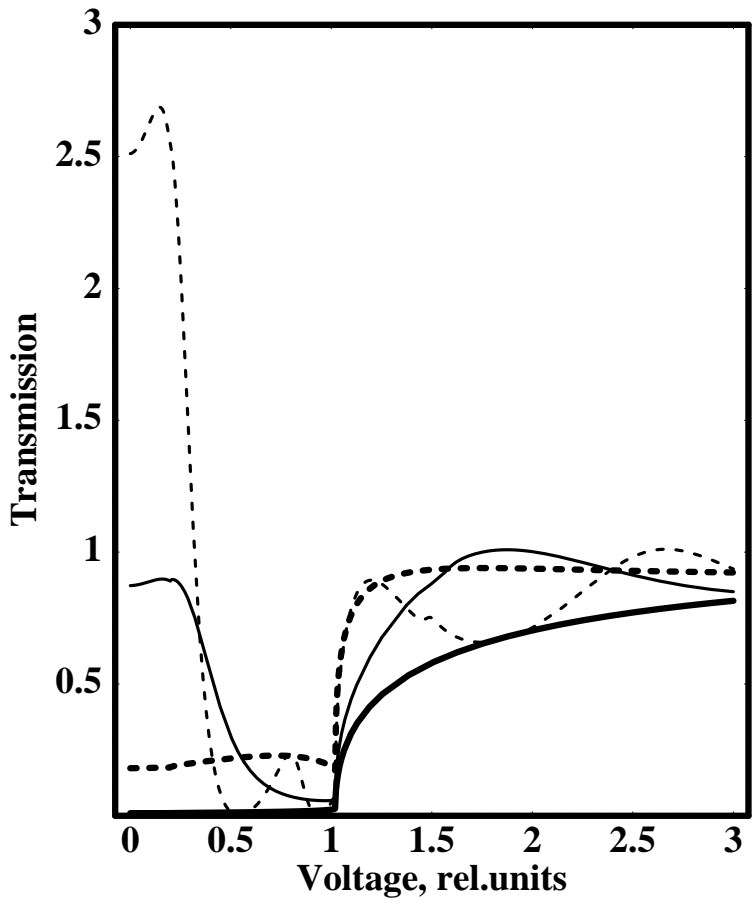
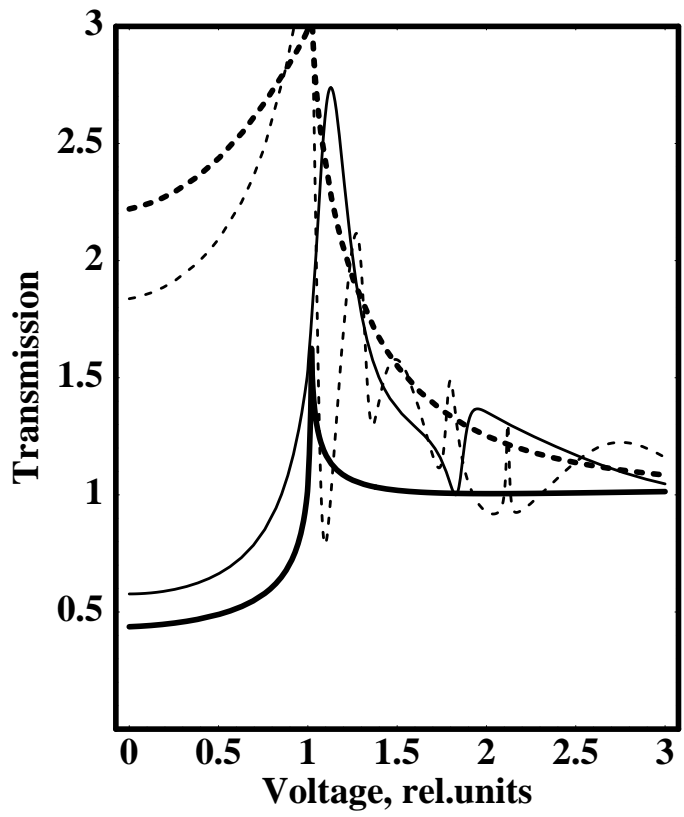
\end{document}